\documentclass[12pt]{revtex4}
\usepackage{amsmath}
\usepackage{amssymb}
\usepackage{graphics}

\newcommand\ket[1]{\left|#1\right>}
\newcommand\bra[1]{\left<#1\right|}

\newcommand\Schr{Schr\"odinger\,}

\newcommand\pr{^\prime}

\newcommand\e{\mathrm{e}}

\newcommand\com{_\mathrm{cm}}

\newcommand\ro{\hat\rho}
\newcommand\Ho{\hat H}
\newcommand\xb{\mathbf{x}}
\newcommand\yb{\mathbf{y}}
\newcommand\xbo{\hat{\xb}}
\newcommand\xo{\hat{x}}

\newcommand\kb{\mathbf{k}}
\newcommand\no{\hat n}
 \newcommand\si{\sigma}

\begin{document}
\title{How to teach and think about spontaneous wave function collapse theories: not like before}   
\author{Lajos Di\'osi}
\affiliation{Wigner Research Centre for Physics\\
H-1525 Budapest 114. P.O.Box 49, Hungary}
\date{\today}
\begin{abstract}
A simple and natural introduction to the concept and formalism of
spontaneous wave function collapse can and should be based on textbook knowledge
of standard quantum state collapse and monitoring. This approach explains the origin
of noise driving the paradigmatic stochastic \Schr equations of spontaneous localization
of the wave function $\Psi$. It reveals, on the other hand, that these equations are empirically
redundant and the master equations of the noise-averaged state $\ro$ are the only empirically
testable dynamics in current spontaneous collapse theories.
\end{abstract}
\maketitle

\section{Introduction}\label{I}
\emph{``We are being captured in the old castle of standard quantum mechanics. Sometimes we think that we have
walked into a new wing. It belongs to the old one, however'' \cite{Dio00a}.}
\vskip 6pt                 
The year 1986 marked the birth of two theories, prototypes of what we call theory of spontaneous wave function collapse.
Both the GRW paper published in Physical Review D \cite{GRW86}
followed by Bell's insightful work \cite{Bel87} and the author's thesis \cite{Dio86}
constructed strict stochastic jump equations to explain unconditional emergence of classical behavior in large quantum systems. 
Subsequently, both theories obtained their time-continuous versions, driven by white-noise
rather than by stochastic jumps.
The corresponding refinement of the GRW proposal \cite{Gis89,GhiPeaRim90,GhiGraBen95} is the Continuous Spontaneous 
Localization (CSL) theory, the author's  gravity-related spontaneous collapse theory \cite{Dio86,Dio87,Dio89} 
used to be called DP theory after Penrose concluded to the same equation for the characteristic time of spontaneous collapse 
in large bodies \cite{Pen96}. These theories modified the standard theory of quantum mechanics in order to describe the 
irreversible process of wave function collapse. The mathematical structure of modification surprised  the proponents 
themselves and it looked strange and original for many of the interested as well. In fact, these theories were considered 
new physics with new mathematical structures, to replace standard equations like \Schr's. 
The predicted effects of spontaneous collapses are extreme
small and have thus remained untestable for the lack of experimental technique. After three decades, fortunately, tests on nanomasses
are now becoming gradually available. Theories like GRW, CSL, DP have not changed over the decades apart from their parameter ambiguities, 
see reviews by Bassi \textit{et al.} \cite{BasGhi03,Basetal13}.
But our understanding and teaching spontaneous collapse should be revised radically. 

Personally, I knew that GRW's random jumps looked like unsharp measurements but, in the late 1980s, I believed that
unsharp measurements were phenomenological modifications of von Neumann standard ones. My belief extended
also for the time-continuous limit of unsharp measurements \cite{Dio88a} that DP collapse equations \cite{Dio89} were based on. 
Finally in the 1990s I got rid of my ignorance and learned that unsharp measurements and my time-continuous measurement 
(monitoring)  could have equally been derived from standard quantum theory \cite{Car93,WisMil10}. 

That was disappointing \cite{Dio00a}. 
Excitement about the radical novelty of our modified quantum mechanics evaporated.
Novelty got reduced to the concept that tiny collapses, that get amplified for bulk
degrees of freedom, happen everywhere and without measurement devices. 
That's why we call them spontaneous.
But they are standard collapses otherwise. 
I have accordingly stressed upon their revised interpretation recently \cite{Dio00b},
the present work is arguing further toward such demand.

\section{How to teach GRW spontaneous collapse?}\label{II}
We should build as much as possible on standard knowledge, using standard concepts, equations, terminology. Key notion is unsharp 
generalized measurement, which has been standard ever since von Neumann showed how
inserting an ancilla between object and measuring device will control measurement unsharpness 
\cite{Neu32}. Hence we are in the best pedagogical position to explain GRW theory to educated physicists. 
No doubt, for old generations measurement means the 
projective (sharp) one but this has changed recently due to the boom in
quantum information science. 
For younger scientists, generalized measurements
are the standard ones, projective measurements are the specific case \cite{NieChu00,Dio11}. 
For new generation, there is  a natural way to get acquainted with spontaneous collapse.
The correct and efficient teaching goes like this.  

GRW theory assumes that independent position measurements of unsharpness (precision) $\si=r_C/\sqrt{2}$, 
with  GRW choice $r_C=10^{-5}$ cm, are happening 
randomly at average frequency $\lambda=10^{-19}$ Hz on each (non-relativistic) particle in the Universe. 
The two parameters $r_C,\lambda$ are considered new universal constants of Nature.
The mathematical model of unsharp measurements is exactly the same as for independent von Neumann detectors \cite{Neu32} 
where the Gaussian ancilla wave function has the width $\si=r_C/\sqrt{2}$ \cite{Dio00a}.
The difference from standard von Neumann detection is the concept of being spontaneous: 
GRW measurements are supposed to happen \emph{without} presence of detectors.

The merit of GRW is wave function \emph{localization} in bulk degrees of freedom
like, e.g., the center-of-mass (c.o.m.) of large objects. 
Quantum theory allows for arbitrary large quantum fluctuations of macroscopic degrees of freedom
in large quantized systems. The extreme example is a \Schr cat state where two macroscopically different wave
functions would be superposed. In GRW theory such macroscopic superpositions or fluctuations become suppressed
by GRW spontaneous measurements but the superpositions of microscopic degrees of freedom will invariably survive.   
These complementary features are guaranteed by the chosen values of parameters $\sigma$ and $\lambda$. Due to the extreme low
rate of measurements, individual particles are almost never measured. But among an Avogadro number $(A)$ of constituents
some $N=A\lambda\sim10^{4}$ become spontaneously measured in each second, meaning that their collective variables, 
e.g.: center-of-mass, become measured each second at precision $\sigma/\sqrt{N}\sim10^{-7}$ cm, 
leading to extreme sharp c.o.m. localization on the long run. That's what we expect of spontaneous localization theories.

The \emph{mathematical model} is the following. 
We model the Universe or part of it by a quantized $N$-body system satisfying the \Schr equation
\begin{equation}\label{Sch}
\frac{d\ket{\Psi}}{dt}=-\frac{i}{\hbar}\Ho\ket{\Psi}
\end{equation}
apart from instances of spontaneous position measurements that happen randomly and independently  
at rate $\lambda$ on every constituent. Spontaneous position measurements are standard generalized measurements. Accordingly,
when the k'th coordinate $\xbo_k$ endures
a measurement, the quantum state undergoes the following collapse:
\begin{equation}\label{GRWj}
\ket{\Psi}\Longrightarrow \frac{\sqrt{G(\xb_k-\xbo_k)}\ket{\Psi}}{\Vert\sqrt{G(\xb_k-\xbo_k)}\ket{\Psi}\Vert}.
\end{equation}
The effects of unsharp position measurement take the Gaussian form:
\begin{equation}
G(\xb_k-\xbo_k)=\frac{1}{(2\pi\si^2)^{3/2}}\exp\left(-\frac{(\xb_k-\xbo_k)^2}{2\si^2}\right),
\end{equation}
where $\xb_k$ is the random outcome of the unsharp position measurement on $\xbo_k$, 
and $\si$ sets the scale of unsharpness (precision). 
The probability of the outcomes $\xb_k$ is defined by the standard rule:
\begin{equation}\label{GRWp}
p(\xb_k)=\Vert\sqrt{G(\xb_k-\xbo_k)}\ket{\Psi}\Vert^2.
\end{equation}

We have thus specified the mathematical model of GRW in terms of standard unsharp position measurements
targeting every constituent at rate $\lambda$ and precision $\si$. 
These measurements are \emph{selective} measurements if we assume that the measurement
outcomes $\xb_k$ are accessible. If they are not, we talk about \emph{non-selective}
measurements and  the jump equation \eqref{GRWj} should be averaged over the
outcomes, according to the probability distribution \eqref{GRWp}. The mathematical model of the GRW theory reduces to the
following master equation for the density matrix $\ro$:
\begin{eqnarray}\label{GRWME}
\frac{d\ro}{dt}&=&-\frac{i}{\hbar}[\Ho,\ro]
+\lambda\sum_k\left(\int d\xb_k\sqrt{G(\xb_k-\xbo_k)}\ro\sqrt{G(\xb_k-\xbo_k)}\right)-\lambda\ro\nonumber\\
                           &=&-\frac{i}{\hbar}[\Ho,\ro]
+\lambda\sum_k \mathcal{D}[\xbo_k]\ro.
\end{eqnarray}
The decoherence superoperator is defined by
\begin{equation}\label{GRWDsupop}
\mathcal{D}[\xbo]\ro=\int d\xb\sqrt{G(\xb-\xbo)}\ro\sqrt{G(\xb-\xbo)}-\ro.
\end{equation}
We can analytically calculate it in coordinate representation $\rho(\xb,\xb\pr)$ of the
density matrix. Its contribution on the rhs of the master equation \eqref{GRWME} shows
spatial decoherence, saturating for large separations:
\begin{equation}\label{GRWdec}
\frac{d\rho(\xb,\xb\pr)}{dt}=~~\dots~~-\lambda\sum_k\left(1-\exp\left(-\frac{(\xb_k-\xb_k\pr)^2}{8\si^2}\right)\right)\rho(\xb,\xb\pr),
\end{equation}
where ellipsis stands for the Hamiltonian part.

The \emph{amplification mechanism} is best illustrated in c.o.m. dynamics. 
As we said, for the individual particles the decoherence term remains negligible
 whereas for bulk degrees of freedom, e.g.: the c.o.m.,  it becomes crucial to damp
\Schr cats, as  we desired. Assume, for simplicity, free spatial motion of a many-body object. 
Then the non-selective GRW  equation \eqref{GRWME} yields  the following autonomous equation for the reduced c.o.m. density matrix $\ro\com$: 
\begin{equation}\label{GRWMEcom}
\frac{d\ro\com}{dt}=-\frac{i}{\hbar}[\Ho\com,\ro\com]
+N\lambda\mathcal{D}[\xbo\com]\ro\com.
\end{equation}
As we see, the decoherence term concerning the c.o.m. coordinate has been
amplified by the number $N$ of the constituents \cite{GRW86} ensuring the desired fast decay of macroscopic superpositions:
\begin{equation}\label{deccom}
\frac{d\rho\com(\xb\com,\xb\com\pr)}{dt}
=~~\dots~~-N\lambda\left(1-\exp\left(-\frac{(\xb\com-\xb\com\pr)^2}{8\si^2}\right)\right)\rho(\xb\com,\xb\pr\com).
\end{equation}
In the selective evolution the individual GRW measurements \eqref{GRWj} entangle the c.o.m., rotation, and internal degrees of freedom,
hence $\ket{\Psi\com}$ does not exist in general. It does in a limiting case of rigid many-body motion when the unitary
evolution of $\ket{\Psi\com}$ is interrupted  by spontaneous $\si$-precision measurements of the c.o.m. coordinate $\xo\com$ 
similarly to \eqref{GRWj} just the average rate of the measurements becomes $N\lambda$ \cite{Dio88b} instead of $\lambda$. 

\section{Localization is not testable, but decoherence is}\label{III}
 Standard concept of selective measurement implies that we have access to the measurement outcomes, which are
the values $\xb_k$ in GRW. If they are accessible variables then the stochastic jump
process of the GRW state vector $\ket{\Psi}$ is testable otherwise it is not. If not, then the same spontaneous measurement is
called non-selective and  what is testable is the density operator $\ro$. The stochastic jump process
(\ref{Sch}-\ref{GRWp}) becomes \emph{illusory} and the master equation \eqref{GRWME} contains the whole GRW physics.

This latter sentence holds in GRW where, as a matter of fact, the $\xb_k$'s remain unaccessible. 
Consider the conservative preparation-detection scenario. Assume we prepared a well-defined pure initial state $\ro_0=\ket{\Psi_0}\bra{\Psi_0}$ 
and by  time $t$ later we desire to test it for the presence of GRW collapses \eqref{GRWj}, we perform no test prior to this one. 
 As a matter of fact, the relevant state is $\ro_t$, being the solution of the master equation \eqref{GRWME} which does not know about
GRW collapses but about GRW decoherence. This is equally valid in the particular case of the macroscopic \Schr cat initial state,
 i.e., a superposition of c.o.m. at two distant locations. The c.o.m. GRW master equation \eqref{GRWMEcom} will exhaustively 
predict the results of all  subsequent tests on the c.o.m. (including the results and statistics of possible naked eye observations).

Obviously, inference on stochastic collapse assumes our access to the measurement outcomes. 
In real laboratory quantum measurements it is the detector design and operation that determine
if we have full (or partial) access to the measurement outcomes or we have no access at all.
In the case of GRW collapse, accessibility of outcomes
it is not a matter of postulation. It is useless to postulate that $\xb_k$'s are
accessible without a  prescription of how to access them.
 
\section{Digression: random unitary process indistinguishable from GRW}\label{IV}
Let us consider an alternative to GRW random process where the stochastic non-linear GRW jumps \eqref{GRWj}
are replaced by the following stochastic  unitary jumps:
\begin{equation}\label{Uj}
\ket{\Psi}\Longrightarrow \e^{i\kb\xbo_k}\ket{\Psi},
\end{equation}
corresponding to the transfer of momentum $\hbar\kb$ to the $k$'th constituent.
The probability distribution of momentum transfer is universal, independent of the particle and of the state:
\begin{equation}\label{Up}
p(\kb)=
\frac{1}{(2\pi\si^{-2})^{3/2}}\exp\left(-\frac{\kb^2}{2\si^{-2}}\right).
\end{equation}
The decoherence superoperator acts as
\begin{equation}\label{UDsupop}
\mathcal{D}[\xbo]\ro=\int d\kb p(\kb) \e^{i\kb\xbo}\ro \e^{-i\kb\xbo}-\ro,
\end{equation}
which looks completely different from the GRW structure \eqref{GRWDsupop} but it coincides with it!
Hence the master equation for the \Schr dynamics \eqref{Sch} with the averaged unitary jumps  \eqref{Uj} will be
the master equation \eqref{GRWME} derived earlier for the GRW theory. 
As we argued in Sec. \ref{III}, the GRW theory can only be tested at the level of the density operator,
no experiment could tell us whether the underlying stochastic process of $\ket{\Psi}$ 
was the GRW stochastic localizing process (\ref{Sch}-\ref{GRWp}) or the above stochastic unitary process.

\section{How to think about CSL?}
We could repeat what we said concerning correct and efficient teaching of GRW in Section \ref{II}.
This time the standard discipline of modern physics, relevant to CSL, is time-continuous
quantum measurement (monitoring) which is just the time-continuous limit of unsharp
sequential measurements similar to those underlying GRW in Section \ref{II}. 
Quantum monitoring theory was not yet conceived in 1986 (GRW), it was born
in 1988, at it became widely known in the nineties, to become the standard theory
of quantum monitoring in the laboratory \cite{Car93,WisMil10}.
It played instrumental role for semi-classical gravity's consistent introduction to spontaneous 
collapse theories \cite{TilDio16,TilDio17}.  
Below I utilize the summary of standard Markovian quantum monitoring theory from \cite{TilDio16}. 

So, how should we interpret CSL? It derives from GRW. The discrete sequence of spontaneous unsharp position measurements
is replaced by spontaneous monitoring the spatial number 
distribution of particles \cite{GhiPeaRim90} (or, in a later version, of the spatial mass distribution of particles 
\cite{GhiGraBen95}). Accordingly, CSL introduces the smeared mass distribution
\begin{equation}\label{dens}
\no(\xb)=\sum_k G(\xb-\xbo_k),
\end{equation}
where, this time, the width of the Gaussian is $r_C$. Monitoring yields the measured signal in the form
\begin{equation}\label{CSLs}
n_t(\xb)=\bra{\Psi_t}\no(\xb)\ket{\Psi_t} +\delta n_t(\xb),
\end{equation}
where $\delta n_t(\xb)$ is the signal white-noise still depending on the spatial resolution/correlation of monitoring. 
The CSL signal noise is a spatially uncorrelated white-noise:
\begin{equation}
\mathbb{E}\delta n_t(\xb)\delta n_s(\yb)=\frac{1 }{4\gamma}\delta(\xb-\yb)\delta(t-s).  
\end{equation}
Just like in the case of GRW sequential spontaneous measurements,
the conditional quantum state evolves stochastically, this time according to
the following stochastic \Schr equation, \emph{driven by the signal noise} in the Ito-sense:
\begin{equation}\label{CSLSSE}
\frac{d\ket{\Psi}}{dt}=\left\{-\frac{i}{\hbar}\Ho-
\frac{\gamma}{2}\int d\xb \left(\no(\xb)-\langle\no(\xb)\rangle\right)^2
+4\gamma\int d\xb \left(\no(\xb)-\langle\no(\xb)\rangle\right)\delta n(\xb)\right\}\ket{\Psi}.
\end{equation}
So far we have introduced the equations of selective spontaneous monitoring, assuming that the
signal \eqref{CSLs} is accessible, which won't be the case, similarly to GRW. In non-selective 
monitoring, the CSL physics reduces to the signal-averaged evolution of the conditional state, i.e., to the CSL master equation:
\begin{equation}\label{CSLME}
\frac{d\ro}{dt}=-\frac{i}{\hbar}[\Ho,\ro]-\frac{\gamma}{2}\int d\xb[\no(\xb),[\no(\xb),\ro]].
\end{equation}
(Note $\gamma=(4\pi r_C^2)^{3/2}\lambda$ would ensure the coincidence with
GRW's spatial decoherence rate at the single particle level although CSL defined a slightly different $\gamma$ \cite{GhiPeaRim90}).

The traditional CSL-teaching differs in a single major point: it does not mention
the theory of monitoring. Hence it does not use the notion of signal $n_t(\xb)$, 
the equation \eqref{CSLs} is not part of  it. Instead, CSL's traditional definition
postulates the stochastic \Schr equation:
\begin{equation}\label{CSLSSEw}
\frac{d\ket{\Psi}}{dt}=\left\{-\frac{i}{\hbar}\Ho-
\frac{\gamma}{2}\int d\xb \left(\no(\xb)-\langle\no(\xb)\rangle\right)^2
+\sqrt{\gamma}\int d\xb \left(\no(\xb)-\langle\no(\xb)\rangle\right) w(\xb)\right\}\ket{\Psi},
\end{equation}
which would correspond to the replacement $\delta n_t(\xb)=2\sqrt{\gamma}w_t(\xb)$ had CSL derived it
from our \eqref{CSLSSE}. The traditional CSL dynamics is driven by the spatially uncorrelated standard white-noise, satisfying
\begin{equation}
\mathbb{E}\delta w_t(\xb)\delta w_s(\yb)=\delta(\xb-\yb)\delta(t-s).  
\end{equation}
In CSL narrative (e.g.: \cite{Basetal13}) the origin of the noise field as well as its anti-Hermitian coupling to density $\no(\xb)$
are mentioned among theory elements yet to be justified, still without reference to
the spontaneous monitoring interpretation available already since long enough time.

Needless to repeat arguments from Section III on GRW. All testable predictions
follow from the CSL master equation \eqref{CSLME}, the stochastic \Schr equation \eqref{CSLSSEw}
is \emph{empirically redundant}, collapse in the claimed quantitative sense is an illusion. 
  
\section{Final remarks}
Disregarding that spontaneous collapse theories are rooted in standard quantum
mechanical collapse theories with hidden detectors has had too many drawbacks.

The principal one is the illusion that the quantitative models of spontaneous collapse 
(localization) in their current forms are 
relevant empirically like master equations of spontaneous decoherence are which have already been
under empiric tests due to recent breakthroughs in technologies. This illusion
is surviving despite no proposals having been ever made for a future experiment to 
test underlying localization effects of $\ket{\Psi}$ beyond decoherence of $\ro$; 
all proposals have so far concerned the dynamical features (e.g.: spontaneous decoherence) of the averaged state $\ro$.

Secondary drawbacks concern illusions that teaching and interpretation
of spontaneous collapse necessitate radical departure from standard
quantum theory both conceptually and mathematically. This may have
kept philosophers excited, may have prevented students of learning
the subject faster, physicists of going deeper into their foundational investigations.

Physics research will gradually adapt itself to the option that spontaneous 
collapse fits better to standard quantum knowledge than we thought of it before.
Monitoring theory roots were revealed for DP spontaneous collapse from the
beginning, and have been detailed and exploited for CSL, too, recently in
\cite{TilDio16,TilDio17}.


\begin{thebibliography}{99}
\bibitem{Dio00a} L. Di\'osi, Emergence of classicality: From collapse phenomenologies to hybrid dynamics, {\it Lect. Notes Phys.} {\bf 538}, 243-250 (2000)
\bibitem{GRW86} G. C. Ghirardi, A. Rimini, and T. Weber, Unified dynamics for microscopic and macroscopic systems, {\it Phys. Rev.} \textbf{D 34}, 470-491 (1986)
\bibitem{Bel87} J. S. Bell: Are there quantum jumps? In {\it Schr\"odinger Centenary of a polymath} (Cambridge, Cambridge University Press, 1987) 
\bibitem{Dio86} L. Di\'osi: A quantum-stochastic gravitation model and the reduction of the wavefunction. Thesis, 1986. Available from: http://wigner.mta.hu/~diosi/prints/thesis1986.pdf
\bibitem{Gis89} N. Gisin, Stochastic quantum dynamics and relativity, {\it Helv. Phys. Acta} \textbf{62}, 363-371 (1989)
\bibitem{GhiPeaRim90} G. C. Ghirardi, P. Pearle, and A. Rimini, Markov processes in Hilbert space and continuous spontaneous localization of systems of identical particles, {\it Phys. Rev.} \textbf{A 42}, 78-89 (1990)
\bibitem{GhiGraBen95} G. C.  Ghirardi, R. Grassi, and F. Benatti, Describing the macroscopic world: Closing the circle within the dynamical reduction program, {\it Found. Phys.}
 \textbf{25}, 5-38 (1995)
\bibitem{Dio87} L. Di\'osi, A universal master equation for the gravitational violation of the quantum mechanics, {\it Phys. Lett.} \textbf{120A}, 377-381 (1987)
\bibitem{Dio89} L. Di\'osi, Models for universal reduction of macroscopic quantum fluctuations, {\it Phys. Rev.} \textbf{A 40}, 1165-1174 (1989)
\bibitem{Pen96} R. Penrose, {\it Gen. Rel. Grav.} {\bf 28}, 581 (1996)
\bibitem{BasGhi03} A. Bassi and G. C. Ghirardi, Dynamical reduction models, {\it Phys. Rep.} \textbf{379}, 257-426 (2003)
\bibitem{Basetal13} A. Bassi, K. Lochan, S. Satin, T. P.  Singh, and H.  Ulbricht, Models of wave-function collapse, underlying theories, and experimental tests, {\it Rev. Mod. Phys.} \textbf{85}, 471-527 (2013)
\bibitem{Dio88a} L. Di\'osi, Continuous quantum measurement and Ito-formalism, {\it Phys. Lett.} \textbf{129A}, 419-423 (1988)
\bibitem{Car93} H. Carmichael: {\it An open system approach to quantum optics} (Berlin, Springer, 1993) 
\bibitem{WisMil10} H. M Wiseman and G. J. Milburn:{\it Quantum measurement and control} (Cambridge, Cambridge University Press, 2010) 
\bibitem{Dio00b} L. Di\'osi, Is spontaneous wave function collapse testable at all? {\it  J. Phys. Conf. Ser.} {\bf 626}, 012008-(5) (2015)
\bibitem{Neu32} J. von Neumann: {\it Mat ematische Grundlagen der Quantenmechanik} (Berlin, Julius Springer, 1932)
\bibitem{NieChu00} M. A.  Nielsen and I. L. Chuang: {\it Quantum computing and quantum information} (Cambridge, Cambridge University Press, 2000)
\bibitem{Dio11} L. Di\'osi: {\it Short course in quantum information theory} (Berlin, Springer, 2011)
\bibitem{Dio88b} L. Di\'osi, On the motion of solids in modified quantum mechanics, {\it Eurohys. Lett.} \textbf{8}, 285-290 (1988) 
\bibitem{TilDio16} A. Tilloy and L. Di\'osi, Sourcing semiclassical gravity from spontaneously localized quantum matter, {\it Phys. Rev.} \textbf{D 93}, 024026-(12) (2016).  
\bibitem{TilDio17} A. Tilloy and L. Di\'osi, Principle of least decoherence for Newtonian semi-classical gravity, {\it Phys. Rev.}  \textbf{D96}, 104045-(6) (2017).
\end{thebibliography}
\end{document}